\documentclass{article}
\usepackage[english]{babel}

\usepackage[letterpaper,top=2cm,bottom=2cm,left=3cm,right=3cm,marginparwidth=1.75cm]{geometry}

\usepackage{amsmath}
\usepackage{graphicx}
\usepackage{subcaption}
\usepackage[colorlinks=true, allcolors=blue]{hyperref}
\usepackage{comment}
\usepackage{listings}
\usepackage{svg}

\title{\textbf{The Mediterraneus Protocol}\\
building an SSI native decentralised ecosystem of digital services}
\author{Luca Giorgino and Andrea Vesco \\
LINKS Foundation -- Cybersecurity Research Group -- 10138 Torino, Italy\\
\{luca.giorgino, andrea.vesco\}@linksfoundation.com
}

\begin{document}
\maketitle

\begin{abstract}
This paper presents, for the first time, the \verb+Mediterraneous+ protocol. It is designed to support the development of an Internet of digital services, owned by their creators, and consumed by users by presenting their decentralised digital identity and a proof of service purchase. \verb+Mediterraneous+ is Self-Sovereign Identity (SSI) native, integrating the SSI model at the core of its working principles to overcome the limitations resulting from using pseudonyms and centralised access control of existing Web3 solutions.
\end{abstract}

\section{Introduction}
\label{sec:introduction}
This paper introduces the \verb+Mediterraneous+ protocol as a collection of building blocks for developing secure and decentralised ecosystems where users can offer and purchase any digital service. The adoption of the \verb+Mediterraneous+ protocol provides the ecosystem built upon it with several fundamental principles.

Users promote and provide their services in a \emph{decentralised} fashion. Users have full control over their services and the power to deliver them using the technology of choice. This design avoids the need to adhere to the specifications typically imposed by a central platform. In this sense, a \verb+Mediterraneous+ protocol-based ecosystem is \emph{self-organising}. The overall behavior of the ecosystem is not enforced but it arises from the deliberate interactions among the users without the need of an orchestrating entity.

Users sell access to their services maintaining \emph{ownership} control. Users, acting as providers, can set the access price and define different access options for their services (\emph{e.g.}, one-time, perpetual, time-bound). 
In addition, users can act as consumers and purchase any service and then access it by proving their digital identity and purchase. 
The process of publishing service offerings and facilitating interactions between providers and consumers is mediated by a smart contract architecture.
Providers offer their services and ensure discoverability and access. Discoverability is facilitated by the minting of Non-Fungible Tokens (NFTs)~\cite{eip-721} representing the services, while access is purchased by exchanging the associated ERC-20~\cite{eip-20} access tokens for a predefined amount of native tokens. 

The resulting peer-to-peer ecosystem preserves the \emph{trustless} principle. There is no need for a central entity to ensure trust between the parties to complete transactions and access the intended services. The distributed ledger and smart contracts enable trusted interactions in a completely decentralised fashion.

The \verb+Mediterraneous+ protocol embraces the Self-Sovereign Identity (SSI) model~\cite{w3c-did,w3c-vc} instead of relying on pseudonyms.
The SSI is integrated into the protocol's core to ensure the \verb+Mediterraneous+ protocol natively supports its principles. Users autonomously manage their identities with a \emph{non-custodial} approach. Apart from joining the ecosystem, the on-chain interactions (\emph{i.e.}, publishing a service offering and purchasing access to a service) and off-chain interactions (\emph{i.e.}, accessing and consuming a service) are mediated by a fully decentralised access control in accordance with the SSI model (see Section~\ref{appendix:a} for a primer on SSI). 

Upon adhering to the \verb+Mediterraneous+ protocol, no entity can prevent anyone in the ecosystem from interacting according to the \emph{censorship-resistant} principle.
A malicious user can always deny off-chain access to another user attempting to consume the purchased service, but he does so at the risk of being labelled as untrustworthy. All users contribute to the self-organization of the decentralised ecosystem, in fact, they have the freedom to select their trusted providers.

In addition, using Zero-Knowledge Proof (ZKP) cryptographic techniques, such as anonymous and selective disclosure on VCs, ensures \emph{privacy preservation}. This empowers users to select their desired level of privacy when interacting with others within the ecosystem.

Integrating the SSI model at the core of the \verb+Mediterraneous+ protocol provides a distinctive value proposition. The adoption of the \verb+Mediterraneous+ protocol enables an Internet of Services, owned by their creators and accessed by users presenting their interoperable, privacy-preserving and decentralised digital identity and proof of service purchase.

\section{Related work} 
\label{sec:related-work}
The closest work to the \verb+Mediterraneous+ protocol is the Ocean protocol~\cite{ocean}. While Ocean offers a comprehensive framework for launching new data marketplaces, including tools and sustainable tokenomics principles, the \verb+Mediterraneous+ protocol focuses on the core technologies that enable the trading of digital services in a decentralised ecosystem, embedding the SSI model at its core~\cite{w3c-did, w3c-vc}. The \verb+Mediterraneous+ protocol embraces the SSI principles to (\emph{i}) provide user with a decentralised digital identity with full control over the data they use to build and prove their identity, and to (\emph{ii}) enable decentralised access control all over the ecosystem. 

In the Ocean protocol, users' identities are tied to the Externally Owned Account (EOA)~\cite{eth-eoa} they use to interact with the marketplace; it is just a pseudonym and not a properly formed identity. The Ocean protocol is privacy-preserving in this sense, however, when it comes to access control~\cite{ocean-docs-1} the protocol needs \emph{i}) marketplace-level permissions to control browsing, downloading or publishing of assets, which requires a central entity to control the marketplace using an old-style Role-Based Access Control (RBAC) server, and 
\emph{ii}) asset-level permissions to control access to a specific asset through access \& deny lists stored in a verifiable data registry, which require significant update and storage costs. 
On the other hand, the \verb+Mediterraneous+ protocol leverages VCs to enable flexible and decentralised access control for on-chain and off-chain interactions. The Issuer(s) in the ecosystem issues VCs to the users and maintains a privacy-preserving binding between the user's EOA and the VC in a smart contract to enable on-chain decentralised access control. Off-chain access control to services is decentralised by design and consists of verifying the consumer's Verifiable Presentation (VP) in accordance with the SSI model, in addition to verifying the proof of purchase.

In addition, the Ocean protocol aims to be trustless and censorship-resistant, meaning that users do not need to rely on others to interact in the marketplace, they have full control over their data, and no one is excluded from participating. However, if access control is enforced, users must trust the central server, which weakens its trustless and censorship-resistant properties. 
In contrast, the \verb+Mediterraneous+ protocol embraces trustless and censorship-resistant principles while allowing decentralised access control through the SSI model.

Both Ocean and \verb+Mediterraneous+ protocol use the ERC-20~\cite{eip-20} tokens to sell, purchase, and control access to assets and services.
While both can support various access options %
the key difference lies in the identity verification of the users.
In the Ocean protocol, a consumer does not need to prove his identity to the provider; possession of the associated ERC-20 token ensures access to the asset. 
In the \verb+Mediterraneous+ protocol, a consumer must hold the associated ERC-20 token and prove his identity to access the service.
There is often tension between users who want to control their privacy and service providers who want to know who is accessing their services. The future integration of Zero-Knowledge VCs into the \verb+Mediterraneus+ protocol is intended to give consumers the power to choose the amount of information they are willing to disclose, and providers the power to set the level of information they require before granting access. In this sense, ZKP capabilities give users the freedom to choose and make their own decisions, thus contributing to the self-organisation of the system.  
In a world where everyone on the Internet is trying to collect more data, the Mediterraneus protocol gives users with full control over their own identity. In this sense, the \verb+Mediterraneus+ protocol has more potential for adoption than Ocean, which does not offer this option.

\section{Infrastructure}
\label{sec:infrastructure}
The \verb+Mediterraneus+ protocol runs on top of a decentralised infrastructure made of the following three components:  

\begin{itemize}

    \item \textbf{Verifiable Data Registry} (VDR): is the root of trust for Decentralised IDentifiers (DIDs)~\cite{w3c-did}, public keys and other data critical to the use of VCs. Examples include distributed ledgers, decentralised file systems and other forms of trusted data storage.
    
    \item \textbf{Distributed Data Storage} (DDS): is the repository for any type of data, readily accessible to all users within the ecosystem. Its primary function is to avoid the use of distributed ledgers to store large amounts of information, thereby mitigating the high storage costs and latency associated with writing to such ledgers.
    
    \item \textbf{Smart Contract Platform} (SCP): is a decentralised platform that provides a layer for deploying Smart Contracts (SCs) that ensure trustless computation within the ecosystem. Ethereum~\cite{ethereum} is the most prominent example of an SCP.

\end{itemize}

The \verb+Mediterraneus+ protocol is theoretically functional with any implementation of these three components. 
The current open source implementation~\cite{mediterraneus-docs} uses the IOTA distributed ledger~\cite{iota} as the VDR, the InterPlanetary File System (IPFS)~\cite{ipfs} as the DDS, and EVM-based~\cite{evm} IOTA Smart Contracts~\cite{isc} as the SCP.

These choices lead us to adopt the IOTA Identity~\cite{iota-identity} library for building the \verb+Mediterraneus+ protocol while ensuring compliance with SSI standards \cite{w3c-did, w3c-vc}.

\section{Actors and Entities}
\label{sec:actors}
\begin{figure}[t]
    \centering
    \includegraphics[width=0.4\columnwidth]{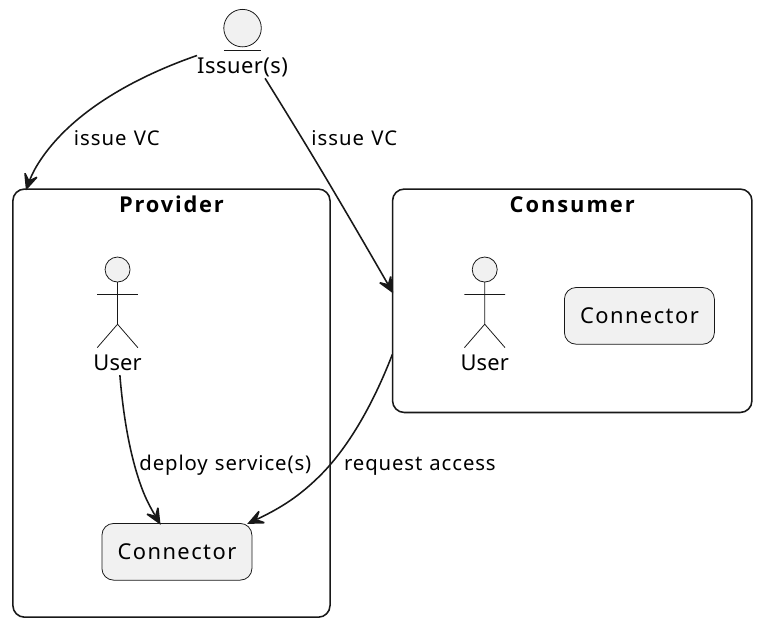}
    \caption{Simplified interactions between entities.}
    \label{fig:actors}
\end{figure}

The actors of the ecosystem, the entities involved in the \verb+Mediterraneus+ protocol and their off-chain interactions are depicted in Figure \ref{fig:actors}.

\begin{itemize}

    \item \textbf{Users} are the actors in the ecosystem that provide and/or consume services. Users interact within the ecosystem through their browsers and connectors. Providers make their services available and accessible through their connectors. Consumers use their wallets to authorise payment transactions to purchase access to services. Providers verify a consumer's identity and purchase before granting access to the requested service.

    \item \textbf{Connector} is the entity delegated by the user to provide persistent and secure storage of identity data and to manage access to services.

    \item \textbf{Issuer(s)} is the entity in charge of issuing VCs to the users willing to join the ecosystem. The Issuer asserts claims about users, creates VCs based on those claims, and issues them to the users. They are also responsible for managing the revocation of the issued VCs. 
    
\end{itemize}

An architecture of SCs, described in Section~\ref{sec:smart-contracts}, mediates interactions between users. They are another important entity that contributes to the realisation of the fundamental principles discussed in Section~\ref{sec:introduction}.

\section{Smart Contract Architecture}
\label{sec:smart-contracts}
\begin{figure}
    \centering
    \includegraphics[width=0.55\columnwidth]{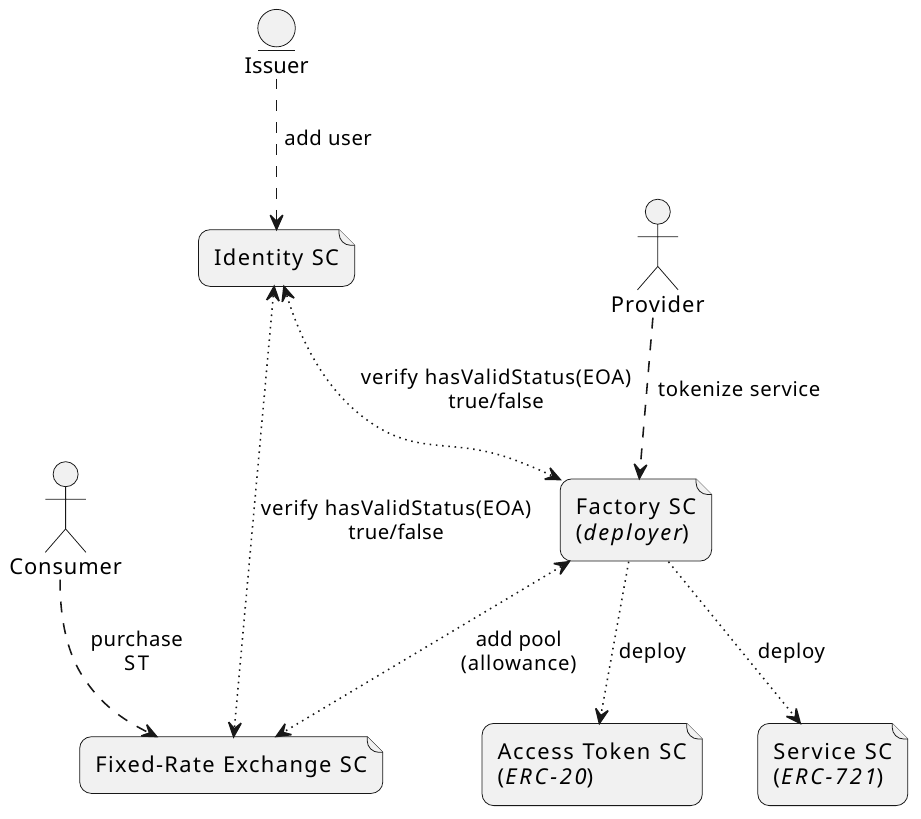}
    \caption{Architecture of smart contracts and interactions between actors (dashed and dotted lines represent transactions issued by actors and SCs respectively on the SCP).}
    \label{fig:smart-contracts}
\end{figure}

The architecture of SCs at the core of the \verb+Mediterraneus+ protocol and the on-chain interactions between actors and SCs are shown in Figure~\ref{fig:smart-contracts}.

\begin{itemize}

\item \textbf{Factory SC} uses the factory pattern to deploy new instances of Service SC and Access Token SC. Users who have joined the ecosystem can interact with the Factory SC to deploy these two contracts, thus minting an NFT and a number of Access Tokens on the SCP (\emph{i.e.} tokenization of a service). 
Upon successful deployment, the Factory SC transfers ownership of the two SC instances to the provider and emits an event signalling successful tokenization. The Factory SC also keeps track of all deployed SC instances to simplify the discovery of tokenized services.

\item \textbf{Service SC} is an ERC-721 compliant contract~\cite{eip-721}, and therefore is an NFT. Each instance of this SC holds the relevant service information and represents the specific tokenized service. Besides default functionalities, this SC exposes the \lstinline{verifyProofOfPurchase()} function. The providers use it to check if a consumer has purchased access to the service. This function internally verifies the balance of the consumer's wallet maintained by the Access Token SC.  

\item \textbf{Access Token SC} is an ERC-20 compliant contract \cite{eip-20} that defines the Access Tokens (ATs). The service provider determines the number of ATs to mint. Consumers purchase the AT to access the service.

\item \textbf{Router SC} maintains the registry of all active exchange contracts and provides the functionality to add or remove exchange contracts. Currently, only the fixed-rate exchange is implemented.

\item \textbf{Fixed-Rate Exchange SC} implements an exchange of ATs for native tokens of the chosen SCP (\emph{e.g.}, Ether, IOTA). When tokenizing a service through the Factory SC, the provider grants allowance for the minted ATs to the Fixed-Rate Exchange SC, enabling their trading at a given fixed price. The consumer interested in the service interacts with the Fixed-Rate Exchange SC to purchase the AT for the pre-defined number of native tokens. Ownership of 1 AT is a prerequisite for accessing the associated service.

\item \textbf{Identity SC} is the key contract to enable decentralised access control all over the ecosystem based on the SSI model. Identity SC is controlled by the Issuer. The Issuer updates the status of the Identity SC by calling the function \lstinline{addUser()} and passing some key metadata of the VC (\emph{i.e.}, \verb+id+, \verb+issuanceDate+ and \verb+expirationDate+)~\cite{w3c-vc} and the EOA of the user. The Identity SC stores these values in the structure:  

\begin{lstlisting}
    struct VCStatus {
        address userEOA;
        uint256 issuanceDate;
        uint256 expirationDate;
        bool revoked;
    }
\end{lstlisting}
and then updates the mapping \lstinline{EOA} $\rightarrow$ \lstinline{id} and \lstinline{id} $\rightarrow$ \lstinline{VCStatus} accordingly (assuming one VC per EOA). The variable \lstinline{revoked}, set to false by default, implements the revocation status list~\cite{w3c-status-list} of VCs directly within the Identity SC.
The SCs that provide on-chain services such as the Factory SC and the Fixed-Rate Exchange SC can call the \lstinline{hasValidStatus()} function passing the value EOA to check if the source of a transaction holds a VC neither expired nor revoked before granting and denying access to their on-chain services, see Figure~\ref{fig:decentalized-access-control} (left).
The entities that provide off-chain services such as the providers' connectors can verify the VP presented by a consumer in accordance with the SSI model while calling the \lstinline{isRevoked()} function, passing the \lstinline{id} of the VC, to check the revocation status, see Figure~\ref{fig:decentalized-access-control} (right).

The on-chain identity verification is challenging for several reasons: cryptographic operations are expensive, SCPs and VDRs are in principle not interconnected, and VPs and VCs should not be exposed directly on the SCP through transactions with the SCs for security and privacy reasons. The Identity SC overcomes these problems and provides a simple solution for implementing decentralised access control on-chain and off-chain.

\begin{figure}
\centering
\begin{subfigure}{.6\textwidth}
  \centering
    \includegraphics[width=0.90\columnwidth]{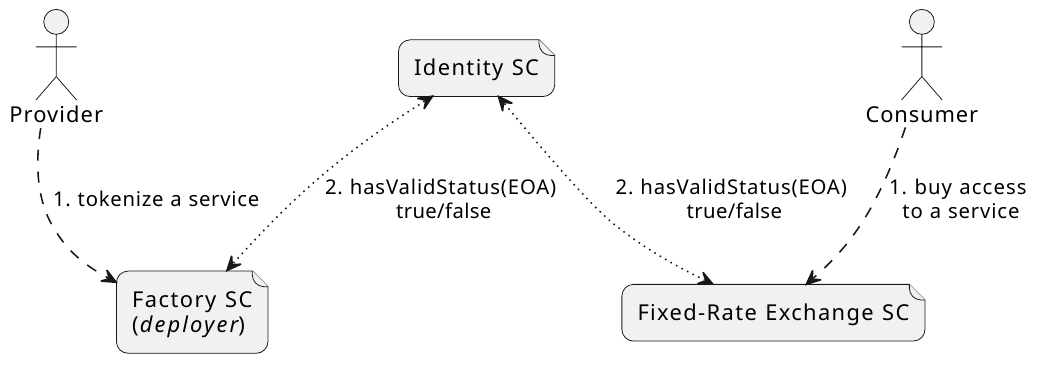}
    \label{fig:identity-sc-1}
\end{subfigure}%
\begin{subfigure}{.4\textwidth}
  \centering
    \includegraphics[width=0.90\columnwidth]{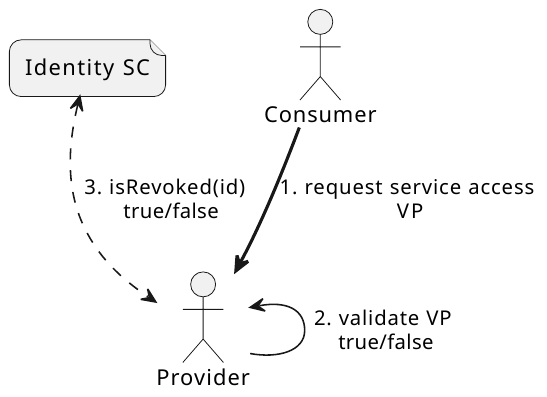}
    \label{fig:identity-sc-2}
\end{subfigure}
\caption{Interactions with the Identity SC to implement distributed access control (dashed and dotted lines represent transactions issued by actors and SCs respectively on the SCP, continuous line represents an off-chain interaction).}
\label{fig:decentalized-access-control}
\end{figure}
\end{itemize}

\section{The Protocol}
\label{sec:protocol}
\subsection{Joining}
\label{sec:protocol-joining}

\begin{figure}[hb]
    \centering
    \includegraphics[width=0.7\columnwidth]{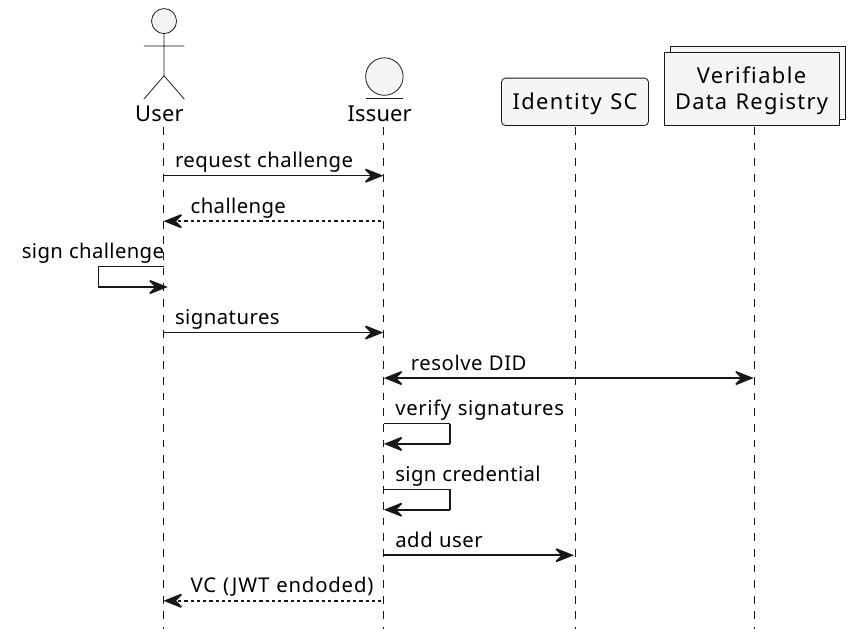}
    \caption{Steps to request a VC from the Issuer.}
    \label{fig:req-cred}
\end{figure}

A user who wants to join the decentralised ecosystem needs a cryptocurrency wallet (w) capable of interacting with the SCP. The wallet has at least an EOA derived from the wallet private key $s_k^w$~\cite{ethereum-book}. In addition, the user must build his own identity in accordance with the SSI model as follows: 

\begin{itemize}
    \item creates the DID, that is
    \begin{itemize}    
        \item generates the identity key pair $(s_k^{id},p_k^{id})$ and stores $s_k^{id}$ securely;
        \item builds the DID Document~\cite{w3c-did} containing two types of verification methods, the \lstinline{JsonWebKey} type to store the identity public key $p_k^{id}$ and the \lstinline{EcdsaSecp256k1RecoveryMethod2020} type to store the EOA
    
        \begin{lstlisting}
        {   
            "id": "did:did-method-name:method-specific-id",
            "verificationMethod": [{
                ...
                "type": "JsonWebKey",
                "publicKeyJwk": "identity public key"
            },
            {   
                ...
                "type": "EcdsaSecp256k1RecoveryMethod2020",
                "blockchainAccountId": "EOA"
            }]
        };
        \end{lstlisting}
        
        \item publishes the DID document in the VDR using the specific DID Method;
    \end{itemize} 
    \item requests a VC to the Issuer following the steps illustrated in Figure~\ref{fig:req-cred}. 
\end{itemize}

The user demonstrates ownership and control of both the DID and EOA to the Issuer by signing a random challenge provided by the Issuer. The user signs the challenge with $s_k^w$ and $s_k^{id}$ generating:
 
$$\sigma_{id} = Sign(s_k^{id}, challenge)$$ 
$$\sigma_w = Sign(s_k^w, challenge)$$

The Issuer resolves the user's DID to a DID Document and extracts $p_k^{id}$ and EOA to proceed with:

$$Verify(p_k^{id}, challenge,\sigma_{id})$$
$$Verify(EOA, challenge,\sigma_w)$$

where the latter checks that (\emph{i}) the signature $\sigma_w$ was generated with $s_k^w$ associated with the EOA~\cite{eip-191}, and (\emph{ii}) the EOA matches the one in the DID document. Upon successful verification of both signatures, the Issuer signs the credential with its private key, updates the status of the Identity SC, and sends the VC back to the user. The VC is JWT encoded.

It is worth noting that the generation of the identity finalized by the issuance of the VC completes the joining procedure of the user.  

\subsection{Publishing}
\label{sec:protocol-publishing}

A provider deploys services on its own connector and makes access to the service available for purchase by other users. The \verb+Mediterraneus+ protocol does not impose any technological choice other than the decentralised access control described in Section~\ref{sec:protocol-accessing}. Once the service is deployed, the provider must upload the description of the service to the DDS, using a data model of their choice, and receive back a URI to the file (\emph{e.g.}, Content ID in case of IPFS).

\begin{figure}
    \centering
    \includegraphics[width=0.7\columnwidth]{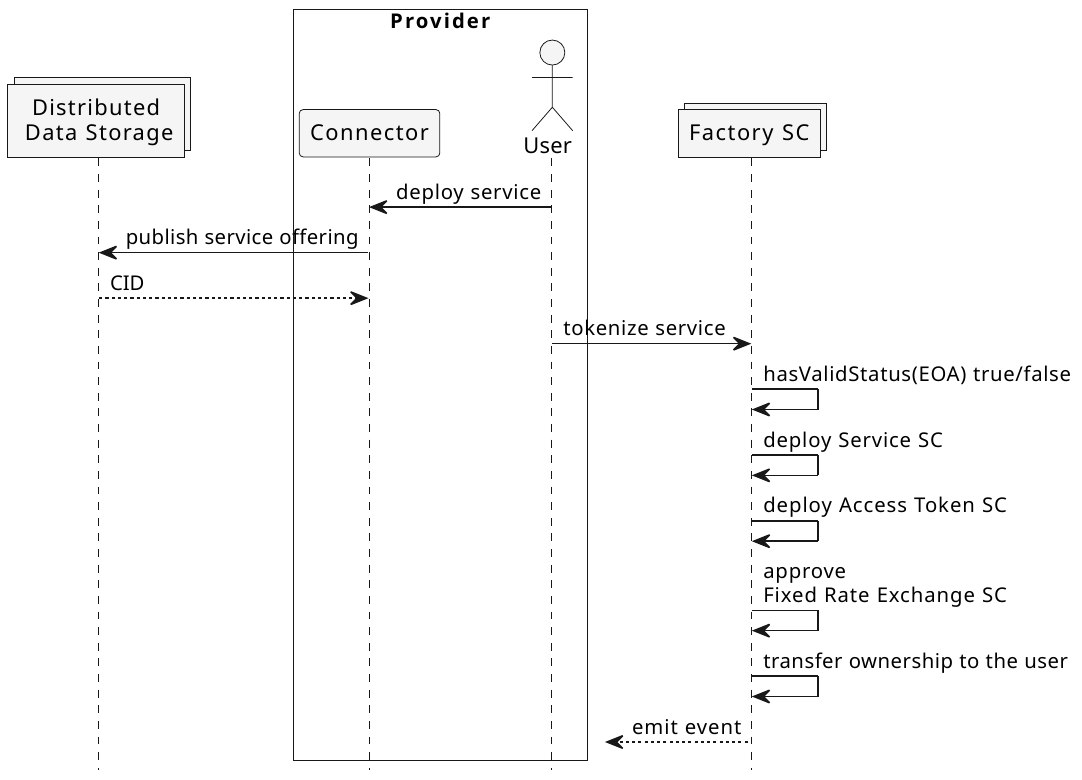}
    \caption{Steps to publish a service offering.}
    \label{fig:pub-asset}
\end{figure}

At this point, the user can publish the service offering by the service tokenization procedure depicted in Figure~\ref{fig:pub-asset}. To tokenize the service the user interacts with the Factory SC to mint the associated NFT and a specified number of ATs on the SCP. 
The service metadata stored in the minted NFT are:
\begin{itemize}

    \item alias: the NFT name chosen by the user;
    
    \item CID: the Content ID to retrieve the file with the description of the service from the DDS;

    \item service URL: the URL for accessing the service.

\end{itemize}
The minting procedure is implemented by the Factory SC as described in Section~\ref{sec:smart-contracts}, in detail the Factory SC:

\begin{itemize}
    \item retrieves the source EOA of the transaction;
    \item interacts with the Identity SC to verify that EOA belongs to a user who has already joined the ecosystem, and the VC is neither expired nor revoked, see Figure~\ref{fig:decentalized-access-control} (left);
    \item if both conditions are met, the Factory SC deploys the new instances of Service SC and Access Token SC, otherwise, it halts the process;
    \item stores the service metadata in the newly minted NFT and transfers ownership of both SCs to the provider;
    \item emits the event of successful tokenization.
\end{itemize}

Upon deployment of the Access Token SC, the specified number of ATs is minted and approved to the Fixed-Rate Exchange SC. This operation activates the sale of the service since it allows the Fixed-Rate Exchange SC to exchange an AT for the predefined amount of native tokens on behalf of the provider. At any time, the provider that is the owner of the Access Token SC can invoke the contract to mint or burn a number of ATs.

\subsection{Purchasing}
\label{sec:protocol-purchasing}
\begin{figure}
    \centering
    \includegraphics[width=0.75\columnwidth]{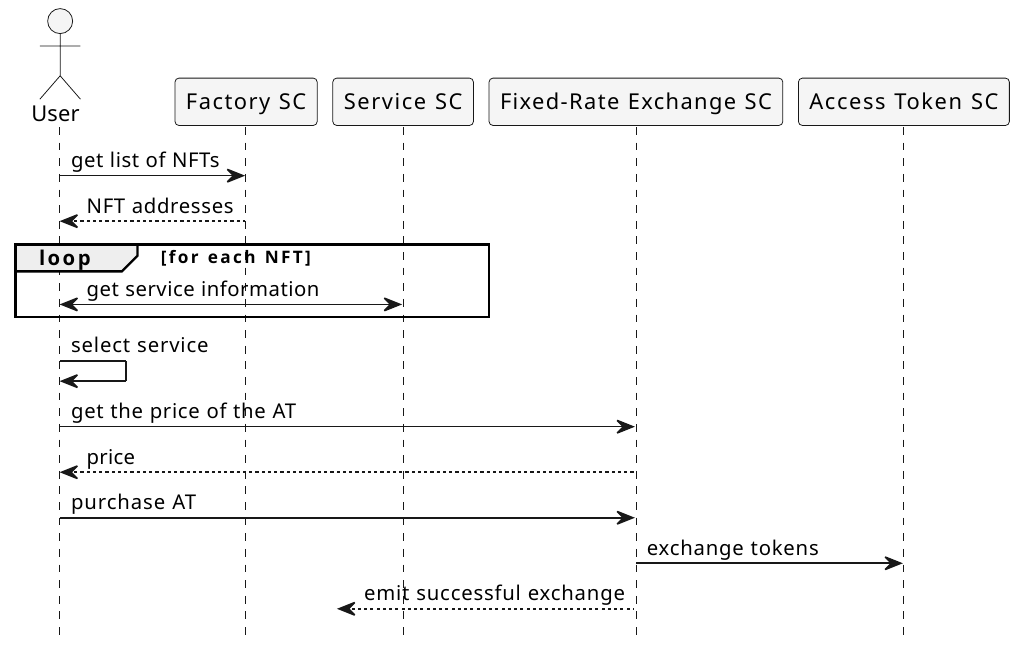}
    \caption{Steps to purchase an AT for the predefined amount in native tokens.}
    \label{fig:purchase-asset}
\end{figure}

Users have two options to discover available services: interact with (\emph{i}) the Factory SC and receive an up to date list of all minted NFTs, that also contains the URI to the description of the service on the DDS, and/or (\emph{ii}) a dedicated catalogue that offers advance search functionalities. The specification of the catalogue is out of the scope of this paper.
After selecting the service, users interact with the Fixed-Rate Exchange SC to purchase access to the service by exchanging the corresponding AT for the predefined amount in native tokens as shown in Figure~\ref{fig:purchase-asset}. 
The Fixed-Rate Exchange SC checks that the exchange request contains the right amount in native tokens and then verifies that both consumer and provider hold a VC neither expired nor revoked. The underlying idea is to make an exchange only if both actors have a valid identity at that precise moment in time.
The Fixed-Rate Exchange SC initiates the transfer of native tokens from the consumer's wallet to the provider's wallet. Simultaneously, the requested AT is transferred from the provider wallet to the consumer wallet. The overall exchange is executed atomically, ensuring that in case of a revert, the native tokens are returned to the consumer and the AT remains in the provider wallet, unaffected.

\subsection{Accessing}
\label{sec:protocol-accessing}

\begin{figure}[hb]
    \centering
    \includegraphics[width=0.65\columnwidth]{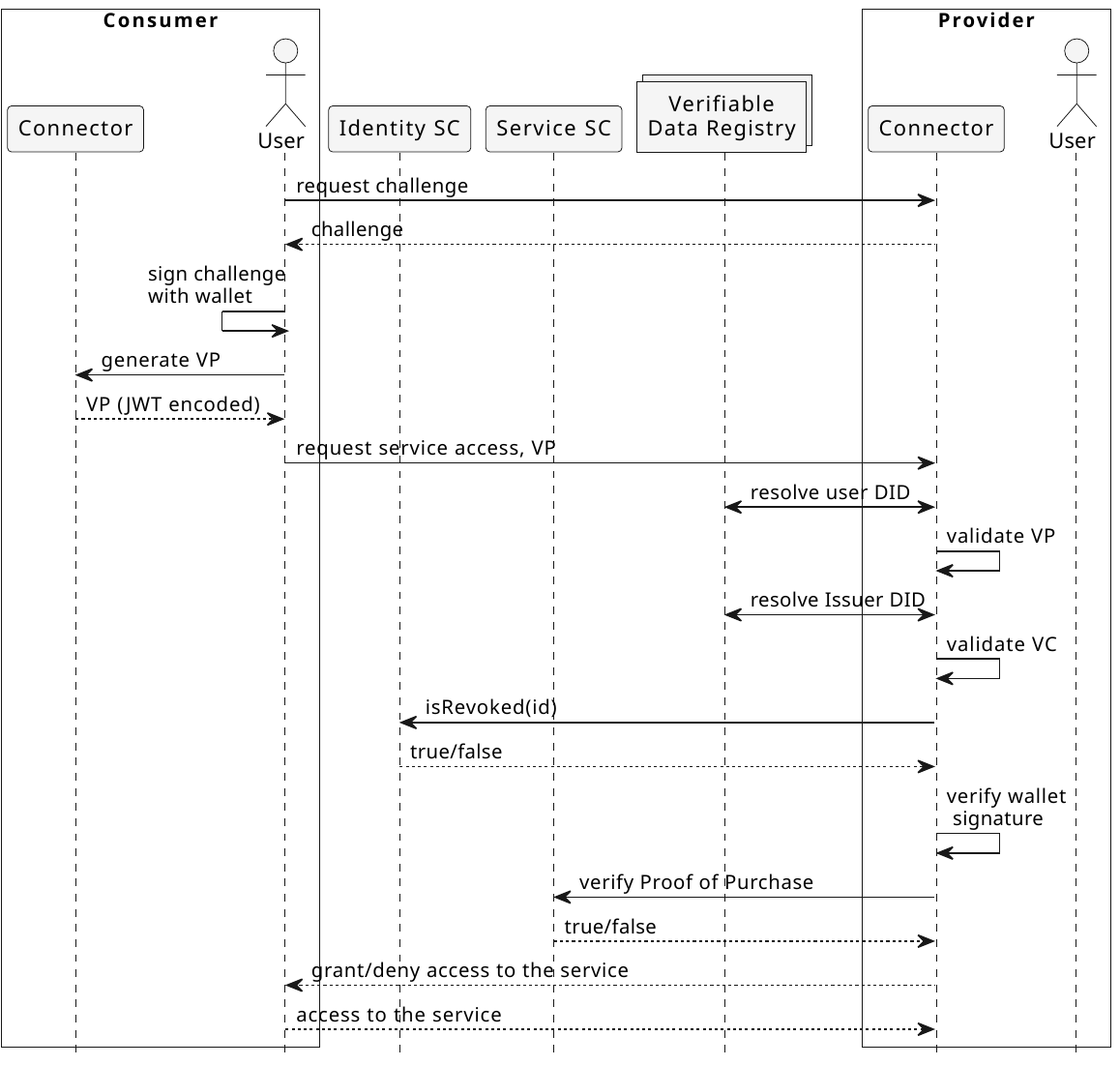}
    \caption{Steps to access a service.}
    \label{fig:access-service}
\end{figure}

The consumer retrieves the service URL from the corresponding NFT and interacts off-chain with the provider's connector to request access to the service. 
Figure~\ref{fig:access-service} shows the steps a consumer follows to access the service: 

\begin{itemize}
 \item requests a random challenge from the provider's connector and signs it with his wallet generating $\sigma_a = Sign(s_k^w, challenge)$;
 \item generates a JWT encoded VP that contains the VC, the challenge, and $\sigma_a$

 \begin{lstlisting}
------------------------ JWT header -----------------------
{ 
  "kid": "did:did-method-name:method-specific-id#fragment",
  "nonce": "... challenge ...",
  ...
}
------------------------ JWT Payload ----------------------
{ ...
  "vp": {
    "type": "VerifiablePresentation", 
    "VerifiableCredential": [ ... ]
  },
  "walletSignature": "... signature sigma_a ..."
}
------------------------ signature ------------------------
     ... user signature with the identity private key ...
\end{lstlisting}

 \item requests access from the provider's connector by presenting the VP.  
\end{itemize}

\noindent Upon receiving a request, the connector performs the following checks to grant or deny access:

\begin{itemize}

    \item extracts the JWT encoded VP from the request;
    \item extracts the consumer's DID from the inner VC and resolves it to the DID Document;
    \item validates the VP with the expected challenge;
    \item extracts the Issuer's DID and resolves it to the DID Document;
    \item validates the VC also checking the VC revocation status as in Figure~\ref{fig:decentalized-access-control} (right);
    \item if both VP and VC are valid, it extracts the signature $\sigma_a$ from the VP and proceeds to $$Verify(EOA, challenge,\sigma_a)$$ that is checks that (\emph{i}) the signature $\sigma_a$ has been generated with $s_k^w$ associated with the EOA~\cite{eip-191}, and (\emph{ii}) the EOA matches the one in the DID document;
    \item interacts with the Service SC to check that the consumer has purchased the related AT as described in Section~\ref{sec:smart-contracts}; 
    \item if the VP verification and the proof of purchase are successful, the provider's connector grants access to the consumer, otherwise denies it.

\end{itemize}

\section{Conclusions and Future Works}
\label{sec:conclusions}
This paper has presented the working principles of the \verb+Mediterraneous+ protocol and has discussed its fundamental principles. The protocol is designed to enable SSI-based decentralised access control, thus fully respecting and supporting the decentralised, trustless, censorship-resistant, self-organizing and non-custodial principles envisaged for the Internet of Services of the future. The Identity SC underpins the current design and implementation of the \verb+Mediterraneous+ protocol. 

Future work in the medium term will focus on (\emph{i}) integrating the ZK VCs, based on the BBS+ signature~\cite{irtf-cfrg-bbs-signatures-05}, to provide users with the ability to present anonymous and selective disclosure VC for privacy preservation while limiting linkability across the ecosystem, (\emph{ii}) updating the working principles of the protocol to accept VCs issued by different issuers, exploiting the natural and very important interoperability feature of the SSI model; while this is straightforward for off-chain interactions, it is not straightforward for on-chain interactions, and (\emph{iii}) implementing different service access options (\emph{e.g.}, one-time, perpetual, time-bound). In the long term, future research will focus on developing cryptographic techniques for the generation and on-chain verification identity proofs.

\section*{\centering Acknowledgement}
    The Authors want to thank Davide Scovotto for his valuable contribution to the design and development of this protocol. 

\bibliographystyle{ieeetr}
\bibliography{paper}

\begin{thebibliography}{10}

\bibitem{web3}
{Ethereum Foundation}, ``{Introduction to Web3}.'' \url{https://ethereum.org/en/web3/}, 2023.

\bibitem{eip-721}
W.~Entriken, D.~Shirley, J.~Evans, and N.~Sachs, ``{ERC-721: Non-Fungible Token Standard, Ethereum Improvement Proposals, no. 721}.'' \url{https://eips.ethereum.org/EIPS/eip-721}, 2018.

\bibitem{eip-20}
F.~Vogelsteller and V.~Buterin, ``{ERC-20: Token Standard, Ethereum Improvement Proposals, no. 20}.'' \url{https://eips.ethereum.org/EIPS/eip-20}, 2015.

\bibitem{w3c-did}
W3C, ``{Decentralized Identifiers (DIDs) v1.0. W3C Recommendation.}.'' \url{https://www.w3.org/TR/did-core/}, 2022.

\bibitem{w3c-vc}
W3C, ``{Verifiable Credentials Data Model v2.0. W3C Working Draft.}.'' \url{https://www.w3.org/TR/vc-data-model-2.0/}, 2023.

\bibitem{ocean}
{Ocean Protocol Foundation}, ``{Ocean Protocol: Tools for the Web3 Data Economy}.'' \url{https://oceanprotocol.com/tech-whitepaper.pdf}, 2022.

\bibitem{eth-eoa}
{Ethereum Foundation}, ``{Ethereum Accounts}.'' \url{https://ethereum.org/developers/docs/accounts}, 2023.

\bibitem{ocean-docs-1}
{Ocean Protocol Foundation}, ``{Fine-Grained Permissions}.'' \url{https://docs.oceanprotocol.com/developers/fg-permissions}, 2023.

\bibitem{ethereum}
{Ethereum Foundation}, ``{Ethereum}.'' \url{https://ethereum.org/en/}.

\bibitem{mediterraneus-docs}
{Fondazione LINKS}, ``{Mediterraneus protocol}.'' \url{https://cybersecurity-links.github.io/mediterraneus}.

\bibitem{iota}
S.~Popov, ``{The Tangle}.'' \url{https://assets.ctfassets.net/r1dr6vzfxhev/2t4uxvsIqk0EUau6g2sw0g/45eae33637ca92f85dd9f4a3a218e1ec/iota1_4_3.pdf}, 2018.

\bibitem{ipfs}
J.~Benet, ``{IPFS - Content Addressed, Versioned, P2P File System. White Paper.}.'' \url{https://github.com/ipfs/papers/raw/master/ipfs-cap2pfs/ipfs-p2p-file-system.pdf}.

\bibitem{evm}
{Ethereum Foundation}, ``{Ethereum Virtual Machine (EVM)}.'' \url{https://ethereum.org/en/developers/docs/evm/}, 2023.

\bibitem{isc}
{Evaldas Drasutis}, ``{IOTA Smart Contracts}.'' \url{https://files.iota.org/papers/ISC_WP_Nov_10_2021.pdf}, 2021.

\bibitem{iota-identity}
{IOTA Foundation}, ``{Digital Identity Framework}.'' \url{https://github.com/iotaledger/identity.rs}.

\bibitem{w3c-status-list}
W3C, ``{Verifiable Credentials Status List v2021: Privacy-preserving status information for Verifiable Credentials. W3C Working Draft.}.'' \url{https://www.w3.org/TR/vc-status-list/}, 2023.

\bibitem{ethereum-book}
G.~W. Andreas M.~Antonopoulos, ``{Mastering Ethereum}.'' \url{https://github.com/ethereumbook/ethereumbook/tree/develop}.

\bibitem{eip-191}
N.~J. Martin Holst~Swende, ``{ERC-191: Signed Data Standard, Ethereum Improvement Proposals, no. 191}.'' \url{https://eips.ethereum.org/EIPS/eip-191}, 2016.

\bibitem{irtf-cfrg-bbs-signatures-05}
T.~Looker, V.~Kalos, A.~Whitehead, and M.~Lodder, ``{The BBS Signature Scheme},'' Internet-Draft draft-irtf-cfrg-bbs-signatures-05, Internet Engineering Task Force, Dec. 2023.
\newblock Work in Progress.

\bibitem{ssi}
A.~Preukschat and D.~Reed, ``{Self-Sovereign Identity: Decentralized digital identity and verifiable credentials}.'' \url{https://www.manning.com/books/self-sovereign-identity}, 2021.

\bibitem{dlt}
N.~Kannengießer, S.~Lins, T.~Dehling, and A.~Sunyaev, ``{Trade-offs between distributed ledger technology characteristics},'' {\em {ACM Computing Surveys}}, vol.~53, no.~2, pp.~1--37, 2020.

\end{thebibliography}

\clearpage
\appendix

\section{Self-Sovereign Identity (SSI)}
\label{appendix:a}
\begin{figure}[t]
    \centerline{\includegraphics[width=\columnwidth]{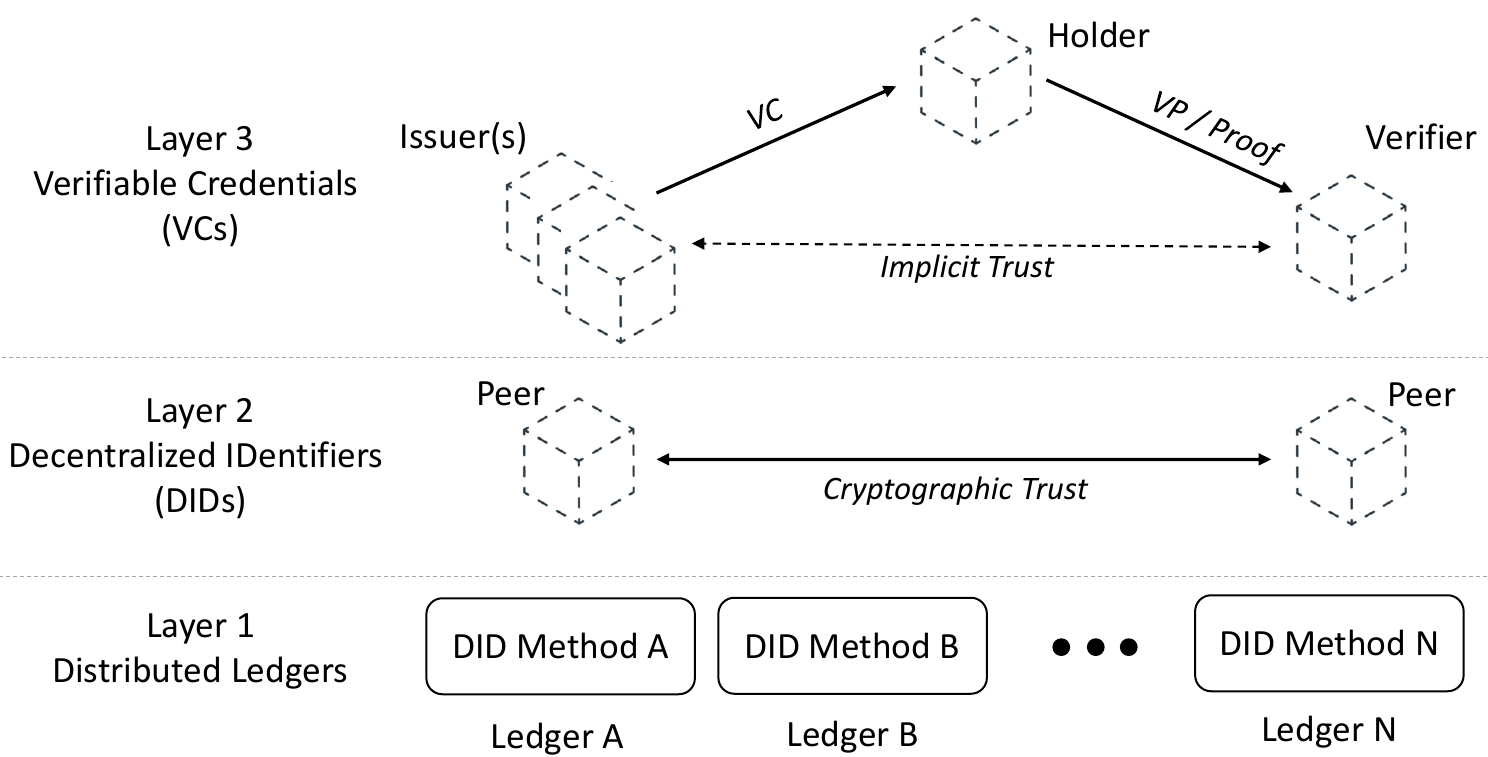}}
    \caption{The Self-Sovereign Identity stack.} 
    \label{fig:ssi}
\end{figure}

The Self-Sovereign Identity (SSI)~\cite{ssi} is a decentralised digital identity model that gives a peer full control over the data it uses to build and prove its identity. The full SSI stack, depicted in Figure~\ref{fig:ssi}, enables a model for trusted interactions.

Layer 1 is implemented by means of any distributed ledger acting as the root of trust for public keys. In fact, distributed ledgers are distributed and immutable means of storage by design~\cite{dlt}. A Decentralized IDentifier (DID)~\cite{w3c-did} is a new type of globally unique identifier designed to verify a peer. The DID is a Uniform Resource Identifier (URI) of the following form:
\begin{center}
    \verb+did:method-name:method-specific-id+
\end{center}
where \verb+method-name+ is the name of the DID Method used to interact with the distributed ledger and \verb+method-specific-id+ is the pointer to the DID Document in the distributed ledger.

Thus, DIDs associate a peer with a DID Document~\cite{w3c-did} to enable trustable interactions with it. The DID Method~\cite{w3c-did} is the software implementation used by a peer to interact with the distributed ledger. A DID Method implementation provides a peer the operations to:
\begin{itemize} 
    \item \textbf{Create} DID: generate an identity key pair ($sk,pk$) for authentication purposes, the corresponding DID Document containing the public key $pk$ and store the DID Document in the distributed ledger at the \emph{method-specific-id} pointed to by the DID; 
    \item \textbf{Resolve} DID: retrieve the DID Document from the \emph{method-specific-id} on the ledger pointed to by the DID; 
    \item \textbf{Update} DID: generate a new key pair ($sk', pk'$) and store a new DID Document to the same or a new \emph{method-specific-id} if the node needs to change the DID; 
    \item \textbf{Deactivate} DID: provide an immutable evidence in the distributed ledger that the DID has been revoked by the owner. 
\end{itemize}
The implementation of the DID method is ledger specific, making the upper layers of the stack independent of the DLT of choice. 

Layer 2 uses DIDs and DID Documents to establish a cryptographic trust between two nodes. In principle, both nodes prove the ownership of their private key $sk$ bound to the public key $pk$ in their DID Document stored in the distributed ledger.

While Layer 2 uses DID technology (i.e. the security foundation of the SSI framework) to start authentication, Layer 3 completes it and also deals with authorisation to services using Verifiable Credentials (VCs)~\cite{w3c-vc}.
A VC is a digital credential that contains additional characteristics of a node's identity beyond its identity key pair, the DID and the DID Document. 
\\

The composition of the identity key pair, the DID, and at least one VC forms the digital identity in the SSI model. The composition of the identity reflects the decentralized nature of SSI. There is no authority that provides all the components of the identity to a peer, and no authority is able to revoke completely the identity of a peer. Moreover, a peer can enrich its identity with multiple VCs issued by different Issuers.
\\
 
Layer 3 works in accordance with the triangle of trust depicted in Figure~\ref{fig:ssi}. Three different roles coexist:
\begin{itemize}
  \item \textbf{Holder} is the peer that owns one or more VCs and generates a Verifiable Presentation (VP) to request services from a Verifier;
  \item \textbf{Issuer} is the peer that asserts claims about the identity of a node, creates a VC from those claims, and signs it before issuing the VC to the Holder, Issuer also deals with revocation of VCs maintaining the revocation status list~\cite{w3c-vc};
  \item \textbf{Verifier} is the node that receives a VP from the Holder and verifies two signatures, one made by the Issuer on the VC and one computed by the Holder on the VP, before granting or denying the access to a service or a resource based on the claims in the VC.
\end{itemize}

The VC contains the metadata to describe the properties of the credential (\emph{e.g.}, \verb+context+, \verb+id+, \verb+type+, \verb+issuer+, and \verb+issuance+ and \verb+expiration+ dates) and most importantly, the DID and the claims about the identity of the peer in the \verb+credentialSubject+ field. The Holder builds the VP as an envelope of the VC, includes the challenge from the Verifier and signs with its private key $s_k$ before presenting it.
\\

It is possible to build any decentralised ecosystem of trustable interactions among peers on top of these three layers.

\end{document}